\documentclass[12pt]{article}
\usepackage{cite}
\usepackage{amsmath, amsthm, amssymb,slashed}
\usepackage{ifpdf}
\ifpdf
  \usepackage[pdftex]{graphicx}
  \usepackage{epstopdf}
\else
  \usepackage[dvips]{graphicx}
\fi
\parskip=\baselineskip

\def\16{{\bf 16}}
\def\1{{\bf 1}}
\def\2{{\bf 2}}
\def\3{{\bf 3}}
\def\4{{\bf 4}}
\DeclareMathAlphabet{\mathpzc}{OT1}{pzc}{m}{it}
\def\bar{\overline}
\def\tilde{\widetilde}

\font\teneurm=eurm10 \font\seveneurm=eurm7 \font\fiveeurm=eurm5
\newfam\eurmfam
\textfont\eurmfam=\teneurm \scriptfont\eurmfam=\seveneurm
\scriptscriptfont\eurmfam=\fiveeurm

\font\teneusm=eusm10 \font\seveneusm=eusm7 \font\fiveeusm=eusm5
\newfam\eusmfam
\textfont\eusmfam=\teneusm \scriptfont\eusmfam=\seveneusm
\scriptscriptfont\eusmfam=\fiveeusm

\font\tencmmib=cmmib10 \skewchar\tencmmib='177
\font\sevencmmib=cmmib7 \skewchar\sevencmmib='177
\font\fivecmmib=cmmib5 \skewchar\fivecmmib='177
\newfam\cmmibfam
\textfont\cmmibfam=\tencmmib \scriptfont\cmmibfam=\sevencmmib
\scriptscriptfont\cmmibfam=\fivecmmib

\numberwithin{equation}{section}

\def\bar{\overline}
\begin{document}

\begin{titlepage}
\begin{flushright}
\end{flushright}
\vskip 1.5in
\begin{center}
{\bf\Large{Ricci flow for homogeneous compact models of the universe}}
\vskip 0.5cm
{Istv\'an Ozsv\'ath} \vskip 0.025in {\small{ \textit{Department of Mathematics, The University of Texas at Dallas}\vskip -.4cm
{\textit{800 W Campbell Rd, Richardson, TX 75080}}}}
\vskip.05in
{Engelbert L. Sch\"ucking} \vskip 0.025in {\small{ \textit{Department of Physics, New York University}\vskip -.4cm
{\textit{4 Washington Place, New York, NY 10003}}}}
\vskip.05in
{Chaney C. Lin} \vskip 0.025in {\small{ \textit{Department of Physics, New York University}\vskip -.4cm
{\textit{4 Washington Place, New York, NY 10003}}}}
\end{center}
\vskip 0.5in
\baselineskip 16pt
\date{March 2011}

\begin{abstract}
Using quaternions, we give a concise derivation of the Ricci tensor for homogeneous spaces with topology of the 3-dimensional sphere. We derive explicit and numerical solutions for the Ricci flow PDE and discuss their properties. In the collapse (or expansion) of these models, the interplay of the various components of the Ricci tensor are studied. We dedicate this paper to honor the work of Josh Goldberg.
\end{abstract}
\end{titlepage}
\section{Prologue}
I learned of Josh's work in relativity first in June 1959. We were invited to participate, and give talks on relativity, at the \emph{Universit\'{e} Libre} in Brussels, the home of the de Donder Condition and George Edward Lema\^{i}tre. This meeting with francophone mathematicians was the prelude to the relativistic oratorio at \emph{L'abbaye de Royaumont} some 30 km north of Paris. A large number of American relativists had been given a free ride to fly from McGuire Air Force Base in New Jersey to Europe on MATS (Military Air Transportation Service) that had delivered the Berlin Air Bridge. The trip on MATS apparently had also done a lot for their egos: to qualify for this perk they had to be given equivalent military ranks to their civilian position. And which graduate student did not cherish to suddenly become a colonel, or which assistant professor did not like to be instantly promoted to one-star general? However, at that time, I did not yet realize that this massive air lift of American relativists was also the work of Josh directed from Wright-Patterson Air Force Base in Ohio. When I mentioned Josh's work, I actually meant a talk about a paper by Josh Goldberg and Ted Newman given in Brussels.

The paper \emph{On the Measurement of Distance in General Relativity} is published in vol. 114 of the Physical Review. I believe that Ted gave the talk because I seem to remember that it was given by a very forceful speaker in an almost breathless way expressing his thoughts with his hands. The essential idea was the use of the formulae for geodesic deviation about null geodesics. It was this technique that led to the beautiful Goldberg-Sachs Theorem. In Royaumont I remember evening walks in the warm summer of the Ilde de France. At that time I had no idea that in this renaissance of relativity Josh was the Duke de Medici. By supporting researchers in America, Britain and Germany and using the resources of the US Air Force Josh contributed greatly to a new start for Einstein's Theory of Gravitation. We hope that historians of relativity will show us the important role that Josh did play for its development. -E.L.S.

\section{Introduction}\label{S:int}
The proof of the Poincar\'e Conjecture by Grigory Perelman \cite{morgan} has raised the interest in Ricci flows.
They were introduced by Richard Hamilton in 1982 \cite{hamilton,chow,topping} and became the basic instrument for the proof.
The flow can be defined as follows.

Let $\mathcal{M}$ be a smooth closed manifold with a smooth Riemannian metric $g$. 
A Ricci flow is the evolution of the metric $g(t)$ under the PDE
\begin{equation}\label{int.1}
\frac{\partial g}{\partial t} = - 2 Ric(g)
\end{equation}
where $Ric(g)$ is the Ricci curvature tensor.
Of particular interest are the Ricci flows in 3-dimensional manifolds.
The normalized Ricci flow equation
\[
\frac{\partial g}{\partial t} = - 2 Ric(g) + \frac{2}{3}\langle R \rangle g,
\]
where $\langle R \rangle$ denotes the average of the scalar curvature $R$ over
the compact 3-manifold, was discussed by James Isenberg and Martin Jackson for
locally homogeneous geometries on closed manifolds \cite{isenberg,knopf}.
We are studying Hamilton's \emph{evolution equation} (\ref{int.1}) which leads
to a collapse of the manifold in a finite timespan.

Over the years, the first two authors have been interested in models of the universe,
and a dozen years ago, they discussed a study on the embedding of compact 3-manifolds into Eucidean spaces in \cite{paper1}.
There, they gave a geometric classification of anisotropically
but homogeneously stretched 3-dimensional spheres and their curvatures.
Here, we wish to complete and extend this discussion by studying
the evolution of these $S^3$'s under the Ricci flow.
It turns out that the 6 non-linear PDE's (\ref{int.1}) reduce to a
single ODE that can be integrated completely in two particularly
interesting cases.
A simple numerical integration provides a picture of the Ricci flow lines.
To keep this paper entirely self-contained, we reproduce the diagrams of \cite{paper1}
and their legends and give the derivation of the Ricci tensor for the evolution equation (\ref{int.1}).
The simple formulae for the principal curvatures give expressions that
look similar to those occuring in the elementary geometry of triangles, namely
those for in- and ex-circles and for Heron's formula.
This lets us expect that some beautiful solid geometry in the tori
of Clifford parallels remains to be discovered in these harmonious universes.
Here we first study the metric and curvature evolution of a deformed
$S^3$ that can be followed in loving detail.
 
\section{A Deformed $S^3$}\label{S:def}
We define an $S^3$ of radius $R$ by the equation
\begin{equation}\label{def.1}
\bar{\xi}\xi = R^2, \quad \xi \in \mathbb{H}.
\end{equation} 
$\bar{\xi}$ is the conjugate of the quaternion $\xi$.
The vectorial quaternionic differential form $\omega$
\begin{equation}\label{def.2}
\omega = \frac{1}{R}\bar{\xi}\,d\xi = -\frac{1}{R}d\bar{\xi}\,\xi = - \bar{\omega}
\end{equation}
can be written as
\begin{equation}\label{def.3}
\omega \equiv \omega_1\,\underline{i} + \omega_2\,\underline{j} + \omega_3\,\underline{k}
\end{equation}
with the three real differential forms $\omega_r$. The index $r$ runs
from one to three. Left-multiplication with the fixed unit quaternion $\lambda$
gives a left-translation of the $S^3$, mapping the point $\xi$ into
the point $\xi'$,
\begin{equation}\label{def.4}
\xi' = \lambda\,\xi, \quad \bar{\lambda}\lambda = 1, \quad  \lambda \in \mathbb{H}.
\end{equation}
We obtain
\begin{equation}\label{def.5}
\omega' \equiv \frac{1}{R}\overline{\lambda\xi}\,d(\lambda\,\xi) = \frac{1}{R}\bar{\xi}(\bar{\lambda}\lambda)\,d\xi = \omega.
\end{equation}
This tells us that the differential forms $\omega$ are left-invariant.
On the $S^3$, $\omega$ is given by $\bar{\xi}d\xi/R$, and we obtain for its metric
\begin{equation}\label{def.6}
ds^2 = d\bar{\xi} \cdot d\xi = \bar{\omega} \cdot \omega = \omega_1^2 + \omega_2^2 + \omega_3^2.
\end{equation} 
This is the metric induced by the Euclidean $\mathbb{R}^4$.
With positive factors $a,b,c$ we define new real differential forms $\tilde{\omega}_r$ by
\begin{equation}\label{def.7}
\tilde{\omega}_1 = \sqrt{\frac{a}{abc}}\omega_1, \quad \tilde{\omega}_2 = \sqrt{\frac{b}{abc}}\omega_2, \quad \tilde{\omega}_3 = \sqrt{\frac{c}{abc}}\omega_3.
\end{equation}
With these new forms we obtain the new metric $d\tilde{s}^2$ for 
the deformed $S^3$ by putting
\begin{equation}\label{def.8}
d\tilde{s}^2 = \bar{\tilde{\omega}} \cdot \tilde{\omega} = \frac{1}{bc}\omega_1^2 + \frac{1}{ac}\omega_2^2 +  \frac{1}{ab}\omega_3^2. 
\end{equation}
This metric, obtained by stretching the $S^3$ in three orthogonal
directions by factors
\[
\sqrt{\frac{a}{abc}}, \quad \sqrt{\frac{b}{abc}}, \quad \sqrt{\frac{c}{abc}},
\]
respectively, is still homogeneous because the differential forms
$\tilde{\omega}_r$ remain left-invariant. Unless $a=b=c$, the
isotropy of the $S^3$ is lost. These beautiful manifolds
were discovered by Luigi Bianchi in 1897 \cite{bianchi}. He called them spaces
of type IX. We prefer the name \emph{Dantes} since the great 
Italian poet envisioned his universe as an $S^3$ \cite{dante,speiser}.

\section{The Ricci Tensor for Dantes}\label{S:ric}
\begin{figure}[htp]
\centering
\includegraphics[width=140mm]{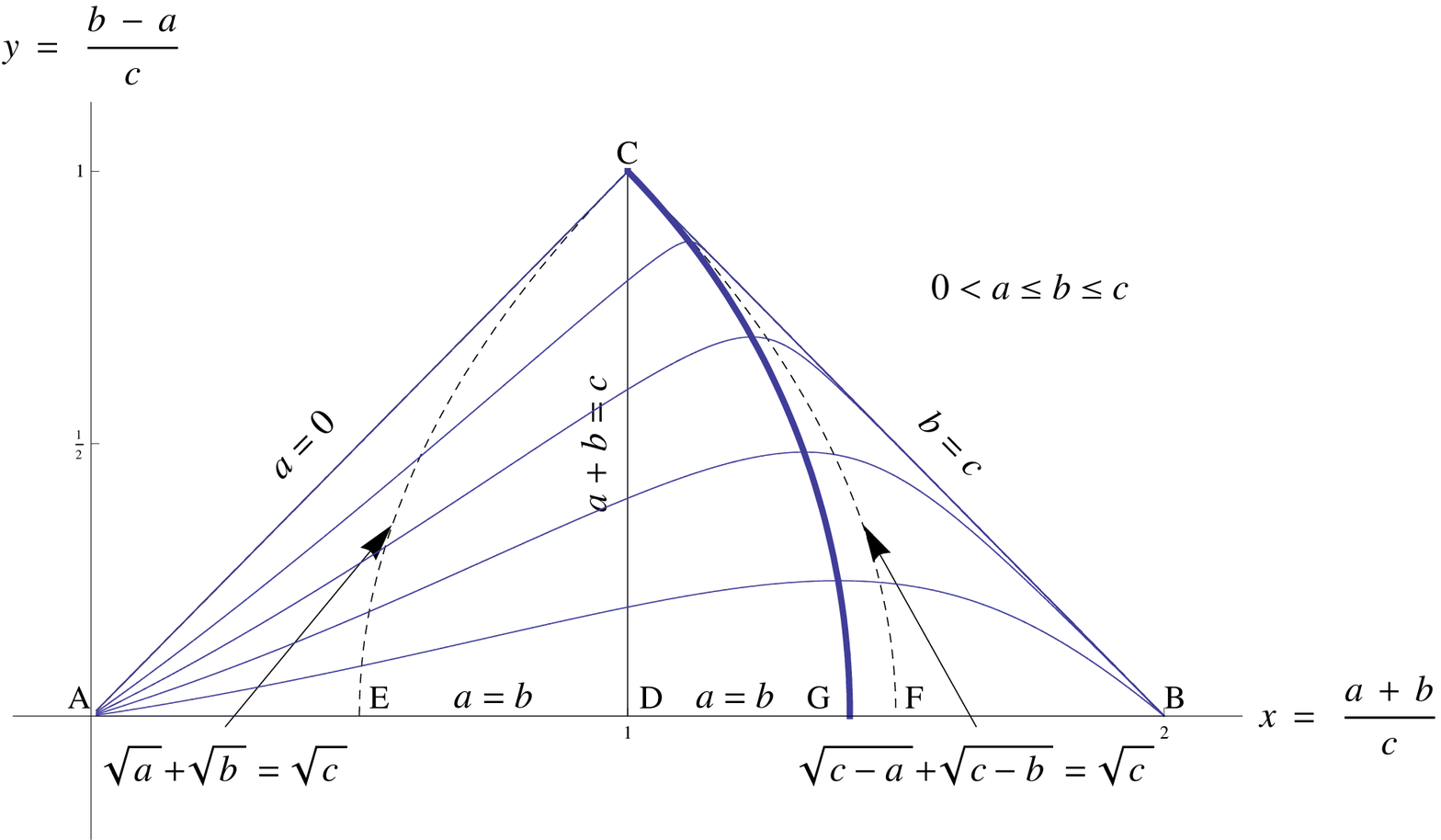}
\caption{
\textbf{Diagram for the stretching ratios of an $S^3$.}
\newline
The $S^3$ metric $ds^2 = \omega_1^2 + \omega_2^2 + \omega_3^3$ with $d\omega_j = -\frac{1}{2}\epsilon_{jkl}\omega_k\wedge\omega_l$ is distorted into $d\tilde{s}^2 = (abc)^{-1}\left[a\omega_1^2+b\omega_2^2+c\omega_3^2\right]$.
\newline
\newline
The hypotenuse of the right-angled triangle $ABC$ covers symmetric snake-like shapes.
The side $BC$ describes turtle-like shapes.
The side $AC$ (with end points included) deals with degenerate configurations which are excluded.
The remaining lines of the Ricci flow connecting points $A$ with $B$ describe dragon-line shapes.
On the bold line $CG$ the flow lines reach their maximum.
The vertex $B$ denotes the isotropically stretched $S^3$. Here $a=b=c$.
\newline
\newline
For the right triangle $BCD$ all eigenvalues of the Ricci tensor are positive.
In its right part $BCF$ all principal curvatures are larger than zero.
On the dashed line $CF$ the smallest principal curvature vanishes,
meaning the largest eigenvalue of the Ricci tensor becomes the sum of the two other eigenvalues.
On the line $CD$ the two smallest eigenvalues of the Ricci tensor take on the value zero.
The principal curvatures are there: $ab, ab, -ab$.
In the left triangle $ACD$ the two smallest eigenvalues of the Ricci tensor are always negative while the largest eigenvalue remains positive.
\newline
\newline
In the region $CBE$ the Ricci scalar, which is twice the sum of the principal curvatures, is positive and vanishes on the dashed line $CE$.
In the domain $ACE$ the Ricci scalar is negative.
}\label{fig.1}
\end{figure}

Differentiation of $\omega$ in (\ref{def.2}) gives with (\ref{def.1}) 
\begin{equation}\label{ric.1}
d\omega = \frac{1}{R}d\bar{\xi} \land d\xi = \frac{1}{R^3}d\bar{\xi} \cdot \xi \land \bar{\xi} \cdot d\xi = - \frac{1}{R}\,\omega \land \omega
\end{equation}
or, written out in full,
\begin{align}\label{ric.2}
0 &= d\omega_1 + \frac{2}{R}\,\omega_2 \land \omega_3, \nonumber\\
0 &= d\omega_2 + \frac{2}{R}\,\omega_3 \land \omega_1,  \\
0 &= d\omega_3 + \frac{2}{R}\,\omega_1 \land \omega_2. \nonumber
\end{align}
These equations for the left-invariant differential forms on the 
$S^3$ are known as the Maurer-Cartan equations. Expressing them in terms of the
new differential forms $\tilde{\omega}_r$ of (\ref{def.7}) we obtain
\begin{align}\label{ric.3}
0 &= d\tilde{\omega}_1 + \frac{2a}{R}\,\tilde{\omega}_2 \land \tilde{\omega}_3, \nonumber  \\
0 &= d\tilde{\omega}_2 + \frac{2b}{R}\,\tilde{\omega}_3 \land \tilde{\omega}_1, \\
0 &= d\tilde{\omega}_3 + \frac{2c}{R}\,\tilde{\omega}_1 \land \tilde{\omega}_2. \nonumber 
\end{align}
Next, we need the connection forms $\tilde{\chi}_r$ ($r = 1,2,3$) for the Dante. For 
vanishing torsion and dimension three, Elie Cartan's first structural equation \cite{kobayashi}
defines $\tilde{\chi}_r$ by the equations
\begin{align}\label{ric.4}
0 &= d\tilde{\omega}_1 + \tilde{\chi}_3 \land \tilde{\omega}_2 - \tilde{\chi}_2 \land \tilde{\omega}_3,  \nonumber \\
0 &= d\tilde{\omega}_2 - \tilde{\chi}_3 \land \tilde{\omega}_1 + \tilde{\chi}_1 \land \tilde{\omega}_3,   \\
0 &= d\tilde{\omega}_3 + \tilde{\chi}_2 \land \tilde{\omega}_1 - \tilde{\chi}_1 \land \tilde{\omega}_2.  \nonumber 
\end{align}
Writing
\begin{equation}\label{ric.5}
\tilde{\chi}_r = a_{rs}\tilde{\omega}_s,
\end{equation}
comparison of (\ref{ric.3}) with (\ref{ric.4}) gives the equations
\begin{align}
&a_{rs} = 0, \quad r \not= s, \\
&a_{22} + a_{33} = -\frac{2a}{R}, \quad
a_{11} + a_{33} = -\frac{2b}{R}, \quad
a_{11} + a_{22} = -\frac{2c}{R}. \nonumber
\end{align}
Addition and subtraction of the equations gives the vectorial connection form
\begin{align}\label{ric.7}
\tilde{\chi}
&= \frac{1}{R}[(a - b - c)\,\tilde{\omega}_1\underline{i} + (b - a - c)\,\tilde{\omega}_2\underline{j} + (c - a - b)\,\tilde{\omega}_3\underline{k}] \\
&= \tilde{\chi}_1\underline{i} + \tilde{\chi}_2\underline{j} + \tilde{\chi}_3\underline{k}. \nonumber
\end{align}
The second structure equation of Cartan determines the curvature
two-forms $\tilde{\Omega}_r$ for dimension three by
\begin{gather}
\begin{pmatrix}
0  &\tilde{\Omega}_3 & -\tilde{\Omega}_2 \\
-\tilde{\Omega}_3 & 0 & \tilde{\Omega}_1 \\
\tilde{\Omega}_2 & -\tilde{\Omega}_1 & 0 
\end{pmatrix}\label{ric.8}\\
=
d
\begin{pmatrix}
0 & \tilde{\chi}_3 & -\tilde{\chi}_2  \\
-\tilde{\chi}_3 & 0 & \tilde{\chi}_1  \\
\tilde{\chi}_2 & -\tilde{\chi}_1 & 0  
\end{pmatrix} 
+ \begin{pmatrix}
0 & \tilde{\chi}_3 & -\tilde{\chi}_2  \\
-\tilde{\chi}_3 & 0 & \tilde{\chi}_1  \\
\tilde{\chi}_2 & -\tilde{\chi}_1 & 0  
\end{pmatrix} \land \begin{pmatrix}
0 & \tilde{\chi}_3 & -\tilde{\chi}_2  \\
-\tilde{\chi}_3 & 0 & \tilde{\chi}_1  \\
\tilde{\chi}_2 & -\tilde{\chi}_1 & 0  
\end{pmatrix}.\nonumber
\end{gather}
By combining the curvature two-forms into a quaternionic vector form
\begin{equation}\label{ric.9}
\tilde{\Omega} = \tilde{\Omega}_1\,\underline{i} + \tilde{\Omega}_2\,\underline{j} + \tilde{\Omega}_3\,\underline{k}
\end{equation}
we can write Cartan's equation (\ref{ric.8})
\begin{equation}\label{ric.10}
\tilde{\Omega} = d\tilde{\chi} - \frac{1}{2}\tilde{\chi} \land \tilde{\chi}.
\end{equation}
This gives with (\ref{ric.7}) and the abbreviation 
\begin{equation}\label{ric.11}
s \equiv \frac{1}{2}(a + b + c)
\end{equation}
the expressions
\begin{align}\label{ric.12}
\tilde{\Omega}_1 &= \frac{4}{R^2}[a(s-a) - (s-b)(s-c)]\,\tilde{\omega}_2 \land \tilde{\omega}_3,  \nonumber \\
\tilde{\Omega}_2 &= \frac{4}{R^2}[b(s-b) - (s-a)(s-c)]\,\tilde{\omega}_3 \land \tilde{\omega}_1,  \\
\tilde{\Omega}_3 &= \frac{4}{R^2}[c(s-c) - (s-a)(s-b)]\,\tilde{\omega}_1 \land \tilde{\omega}_2.  \nonumber 
\end{align}
In three dimensions,
\[
\tilde{\omega}_2 \land \tilde{\omega}_3, \quad \tilde{\omega}_3 \land \tilde{\omega}_1, \quad \tilde{\omega}_1 \land \tilde{\omega}_2
\]
form a basis for all two-forms, and we define
\begin{equation}\label{ric.13}
\tilde{\omega}_2 \land \tilde{\omega}_3 \equiv \tilde{H}_1, \quad \tilde{\omega}_3 \land \tilde{\omega}_1 \equiv \tilde{H}_2, \quad \tilde{\omega}_1 \land \tilde{\omega}_2 \equiv \tilde{H}_3.
\end{equation}
We can then write
\begin{equation}\label{ric.14}
\tilde{\Omega}_r = \sum\limits_{s=1}^3T_{rs}\tilde{H}_s
\end{equation}
where $T_{rs}$ are the symmetric components of the Riemann tensor. The eigenvalues of this tensor are the 
principal curvatures
\[
\kappa_r, \quad r = 1,2,3.
\]
We now can read off from (\ref{ric.12}) the principal curvatures
\begin{align}\label{ric.16}
\kappa_1 &= \frac{4}{R^2}[a(s-a) - (s-b)(s-c)], \nonumber \\
\kappa_2 &= \frac{4}{R^2}[b(s-b) - (s-a)(s-c)],  \\
\kappa_3 &= \frac{4}{R^2}[c(s-c) - (s-a)(s-b)]. \nonumber 
\end{align}
The components $R_{rs}$ of the Ricci tensor are obtained from the components $T_{rs}$
of the Riemann tensor in the three-dimensional case by subtraction of the trace
\begin{equation}\label{ric.17}
- R_{rs} = T_{rs} - \delta_{rs}\sum\limits_{t=1}^3T_{tt},
\end{equation}
where $\delta_{rs}$ is Kronecker's delta. In our case the Ricci
tensor appears also in diagonal form, and we have
\begin{align}\label{ric.18}
R_{11} &= \kappa_2+\kappa_3 = \frac{8}{R^2}(s-b)(s-c), \nonumber \\
R_{22} &= \kappa_1+\kappa_3 = \frac{8}{R^2}(s-a)(s-c), \\
R_{33} &= \kappa_1+\kappa_2 = \frac{8}{R^2}(s-a)(s-b). \nonumber 
\end{align}
The other three independent components of the Ricci tensor vanish.
A classification of the Dantes in terms of their Ricci tensors is
given in Fig. \ref{fig.1}.

\section{Homogeneous Ricci Flow in Dantes}\label{S:hom}
The metric $g(t)$ of the Dantes is given by (\ref{def.8})
\begin{equation}\label{hom.1}
d\tilde{s}^2 = g(t) = \bar{\tilde{\omega}} \cdot \tilde{\omega} = \frac{\bar{\omega}_1 \cdot \omega_1}{b(t)c(t)} + \frac{\bar{\omega}_2 \cdot \omega_2}{a(t)c(t)} + \frac{\bar{\omega}_3 \cdot \omega_3}{a(t)b(t)}. 
\end{equation}
The Ricci tensor of the Dantes is given by
\begin{equation}\label{hom.2}
Ricci = R_{11}\,\bar{\tilde{\omega}}_1 \cdot \tilde{\omega}_1 + R_{22}\,\bar{\tilde{\omega}}_2 \cdot \tilde{\omega}_2 + R_{33}\,\bar{\tilde{\omega}}_3 \cdot \tilde{\omega}_3.
\end{equation}
The equation (\ref{int.1}) defining the Ricci flow becomes
\begin{align}
\dot{g}
&= \Big(\frac{1}{bc}\Big)^.\bar{\omega}_1\cdot\omega_1 + \Big(\frac{1}{ac}\Big)^.\bar{\omega}_2\cdot\omega_2 + \Big(\frac{1}{ab}\Big)^.\bar{\omega}_3\cdot\omega_3 \label{hom.3}\\
&= - 2\frac{R_{11}}{bc}\bar{\omega}_1\cdot\omega_1 - 2\frac{R_{22}}{ac}\bar{\omega}_2\cdot\omega_2 - 2\frac{R_{33}}{ab}\bar{\omega}_3\cdot\omega_3,\nonumber
\end{align}
where we have indicated differentiation with respect to the 
parameter $t$ by a dot. By comparing terms and using (\ref{ric.18})
we obtain the system of three ordinary differential equations
\begin{align}\label{hom.4}
\Big(\frac{1}{bc}\Big)^. &= - \frac{16(s-b)(s-c)}{R^2bc}, \nonumber \\
\Big(\frac{1}{ac}\Big)^. &= - \frac{16(s-a)(s-c)}{R^2ac}, \\
\Big(\frac{1}{ab}\Big)^. &= - \frac{16(s-a)(s-b)}{R^2ab}.\nonumber 
\end{align}
If we introduce the abbreviations 
\begin{equation}\label{hom.5}
u \equiv \frac{1}{bc}, \quad v \equiv \frac{1}{ac}, \quad w \equiv \frac{1}{ab}
\end{equation}
and
\begin{equation}\label{hom.6}
\sigma \equiv \frac{1}{2}(u + v + w)
\end{equation}
we can write the system (\ref{hom.4}) as
\begin{align}\label{hom.7}
\dot{u} &= - \frac{16(\sigma-v)(\sigma-w)}{R^2vw} = - \frac{4}{R^2}\Big[2 + \frac{u^2-v^2-w^2}{vw}\Big], \nonumber  \\
\dot{v} &= - \frac{16(\sigma-u)(\sigma-w)}{R^2uw} = - \frac{4}{R^2}\Big[2 + \frac{v^2-u^2-w^2}{uw}\Big],  \\
\dot{w} &= - \frac{16(\sigma-u)(\sigma-v)}{R^2uv} = - \frac{4}{R^2}\Big[2 + \frac{w^2-u^2-v^2}{uv}\Big]. \nonumber 
\end{align}

\section{The Isotropic Flow}\label{S:iso}
If $a = b = c$, then we have also $u = v = w$, and the system (\ref{hom.7}) becomes
\begin{equation}\label{iso.3}
\dot{u} = - \frac{4}{R^2}.
\end{equation}
This integrates to
\begin{equation}\label{iso.4}
u = \text{const.} - \frac{4t}{R^2}.
\end{equation}
The linear stretch factor is given by
\begin{equation}\label{iso.5}
\lambda = \sqrt{u}.
\end{equation}
If we begin the flow with $\lambda = 1$ at $t = 0$, then the constant in (\ref{iso.4}) becomes 1 and we have
\begin{equation}\label{iso.7}
\lambda = \sqrt{1 - 4t/R^2}.
\end{equation}
This leads to the collapse of an $S^3$ of radius $R$ in
\begin{equation}\label{iso.8}
t = T = R^2/4.
\end{equation}
To simplify the equations in what follows, we shall simply put
\begin{equation}\label{iso.9}
R^2 = 4.
\end{equation}

\section{The Symmetric Flow}\label{S:sym}
The equations for the Ricci flow of Dantes are invariant under a permutation
of the three stretching factors. In the following, we shall assume as an initial
condition for $t = 0$ that
\begin{equation}\label{sym.1}
0 < \sqrt{a_0} \leqq \sqrt{b_0} \leqq \sqrt{c_0}.
\end{equation}
In the previous section, we dealt with the case
\begin{equation}\label{sym.2}
a_0 = b_0 = c_0.
\end{equation}
There, a spherical Dante remains spherical until it collapses into a point.
We wish to show now that the inequalities
\begin{equation}\label{sym.3}
0 < a \leqq b \leqq c
\end{equation}
among the stretching factors of a Dante are preserved during the Ricci flow.

With $R^2 = 4$, we have from (\ref{hom.4})
\begin{align}\label{sym.4}
\Big(\frac{b-a}{abc}\Big)^. &= - 4\Big(\frac{b-a}{abc}\Big)s(s-c),  \\
\Big(\frac{c-b}{abc}\Big)^. &= - 4\Big(\frac{c-b}{abc}\Big)s(s-a). \nonumber
\end{align}
These equations show that $a=b$ and $b=c$ are solutions. Any infinitesimally spherical
solid ball in the $S^3$ will be stretched into a solid infinitesimal triaxial ellipsoid
with principal axes whose lengths are in the ratios $a:b:c$.
We call the case $a=b$ where the two shorter stretching factors are equal the \emph{snake}.
We call the case $b=c$ where the two larger stretching factors are equal the \emph{turtle}.

The equations (\ref{sym.4}) show us  that the Ricci flow preserves the 
equalities and inequalities of (\ref{sym.3}): Suppose we have the 
one-dimensional $y(t)$ such that
\begin{equation}\label{sym.5}
\dot{y}(t) = y(t)f(t)
\end{equation}
with some function $f(t)$ where it is of no interest whether the function is 
positive or negative. The solution is
\begin{equation}\label{sym.6}
y(t) = y_0\exp\int\limits_0^t f(\tau)d\tau, \qquad y_0 = \text{const.}
\end{equation}
The sign of $y_0$ is thus also the sign of $y(t)$.

Fig. \ref{fig.1} shows the Ricci flow in the $x$-direction.
That this coincides with the time direction follows from the fact that
\begin{equation}\label{sna.x1}
\frac{dx}{dt} > 0
\end{equation}
for $0 < x < 2$.
We have with (\ref{hom.5})
\begin{equation}\label{sna.x2}
x = \frac{a+b}{c} = \frac{u+v}{w}
\end{equation}
and with (\ref{hom.7})
\begin{equation}\label{sna.x3}
\dot{x} = \frac{(\dot{u}+\dot{v})w - \dot{w}(u+v)}{w^2}.
\end{equation}
This gives with (\ref{hom.7}) and $R^2 = 4$
\begin{equation}\label{sna.x4}
\dot{x} = \frac{2\left[u(w-v)^2 + u^2(v-u) + v(w^2-v^2)\right]}{uvw^2}.
\end{equation}
Since (\ref{sym.3}) gives with (\ref{hom.5})
\begin{equation}\label{sym.11}
w \geq v \geq u
\end{equation}
we see that $\dot{x}$ vanishes only in the case that
\begin{equation}\label{sym.12}
w = v = u
\end{equation}
that gives the isotropic case for $x=2$.

\section{The Snake}\label{S:sna}
For $a=b$, it follows from (\ref{hom.5}) that $u=v$. The system (\ref{hom.7}) 
reduces to the two differential equations
\begin{equation}\label{sna.1a}
\dot{w} = - \frac{w^2}{v^2}
\end{equation}
and
\begin{equation}\label{sna.1b}
\dot{v} = \frac{w}{v} - 2.
\end{equation}
Multiplying (\ref{sna.1b}) by $v$ and differentiating gives
\begin{equation}\label{sna.2}
\dot{w} = 2 \dot{v} + \dot{v}^2  + \ddot{v}v.
\end{equation}
Substituting into (\ref{sna.1a}) gives a differential equation
for $v$ alone
\begin{equation}\label{sna.3}
 2 \dot{v} + \dot{v}^2  + \ddot{v}v = - (\dot{v} + 2)^2
\end{equation}
or
\begin{equation}\label{sna.4}
\ddot{v}v + 2\dot{v}^2 + 6\dot{v} + 4 = 0.
\end{equation}
It is clear that $\dot{v} = 0$, i.e. constant $v$, cannot be a solution. We do not
lose solutions by taking now $\dot{v}$ as a function of $v$ itself. Introducing 
\begin{equation}\label{sna.5}
\dot{v} \equiv z(v), \quad \ddot{v} = \frac{dz}{dv}\,z,
\end{equation}
the equation (\ref{sna.4}) becomes
\begin{equation}\label{sna.6}
z\frac{dz}{d\log v} = - 2 z^2 - 6 z - 4 = -2(z+1)(z+2).
\end{equation}
Integration gives with constant $v_0$
\begin{equation}\label{sna.7}
\log(\frac{v_0}{v})^2 = \int\frac{z dz}{(z+1)(z+2)} = \int\frac{2dz}{z+2} - \int\frac{dz}{z+1} = \log \frac{(z+2)^2}{z+1}
\end{equation}
or
\begin{equation}\label{sna.8}
(z+2)^2 - (z+1)\Big(\frac{v_0}{v}\Big)^2 = 0.
\end{equation}
Replacing $z = \dot{v}$ by (\ref{sna.1b}) gives the integral
\begin{equation}\label{sna.9}
\Big(\frac{w}{v}\Big)^2 - \Big(\frac{w}{v}-1\Big)\Big(\frac{v_0}{v}\Big)^2 = 0.
\end{equation}
We use now the integral (\ref{sna.9}) to give $w/v$ as a function of $w$. This gives
\begin{equation}\label{sna.10}
\frac{w}{v} = 1 + \Big(\frac{w}{v_0}\Big)^2.
\end{equation}
With (\ref{sna.1a}) we obtain
\begin{equation}\label{sna.11}
\frac{dw}{dt} = - \Big[1 + \Big(\frac{w}{v_0}\Big)^2\Big]^2.
\end{equation}
This integrates to
\begin{equation}\label{sna.13}
\frac{1}{v_0}(t_0 - t) = \frac{w/v_0}{2(1 + (w/v_0)^2)} + \frac{1}{2}\tan^{-1}(w/v_0)
\end{equation}
with constant $t_0$.
For small values of $w/v_0$ we find
\begin{equation}\label{sna.14}
t_0 - t \approx w.
\end{equation}
For large values of $w/v_0$ we find
\begin{equation}\label{sna.15}
\frac{1}{v_0}(t_0 - t) \approx \frac{\pi}{4} + \frac{1}{2w/v_0}.
\end{equation}
With these data we are now able to describe the Ricci flow for the snake. Our 
initial conditions are the values of $w$ and $v$ at time $t=0$. These are the 
squares of the initial stretching factors of the $S^3$. We call these initial values $W$ and $V$.
The ratio $W/V$
\begin{equation}\label{sna.16}
W/V = c(0)/b(0), \qquad a(0) = b(0)
\end{equation}
gives the initial aspect ratio of the snake. We have
\begin{equation}\label{sna.17}
(W/V) - 1 = \alpha^2
\end{equation}
where the parameter $\alpha$ measures the degree of initial non-sphericity. The
integral (\ref{sna.10}) allows us to determine the constant $v_0$ that belongs 
to the given initial values. We get
\begin{equation}\label{sna.18}
v_0 = \frac{W}{\sqrt{(W/V) -1}} = \frac{W}{\alpha}.
\end{equation}
We determine the constant $t_0$ by entering the initial values into 
(\ref{sna.13}) and obtain with (\ref{sna.10})
\begin{align}\label{sna.19}
t_0 &= \frac{W}{2 W/V} + \frac{W}{2\alpha}\tan^{-1}\alpha \nonumber \\
    &= \frac{W}{2}\Big(\frac{1}{1+\alpha^2} + \frac{1}{\alpha}\tan^{-1}\alpha\Big).
\end{align}
We obtain then from (\ref{sna.13}) for the time $t$ as a function of $w$
\begin{equation}\label{sna.21}
t = \frac{W}{2}\left[\frac{1}{1+\alpha^2} + \frac{1}{\alpha}\tan^{-1}\alpha\right] - \frac{1}{2}\left[\frac{w}{1+\alpha^2(w/W)^2} + \frac{W}{\alpha}\tan^{-1}\left(\alpha\frac{w}{W}\right)\right].
\end{equation}
Writing
\begin{equation}\label{sna.22}
\lambda = \frac{w}{W}
\end{equation}
we have
\begin{equation}\label{sna.23}
t = \frac{W}{2}\left[\frac{(1-\lambda)(1-\lambda\alpha^2)}{(1+\alpha^2)(1+\alpha^2\lambda^2)} + \frac{1}{\alpha}\tan^{-1}\frac{(1-\lambda)\alpha}{1+\alpha^2\lambda}\right].
\end{equation}
At $t=0$ we have $\lambda = 1$. As $t$ increases, $\lambda$ shrinks; the snake gets shorter.
The snake collapses to zero length at $\lambda = 0$. We obtain thus for the duration $T$ of the flow
\begin{equation}\label{sna.24}
T = \frac{W}{2}\Big(\frac{1}{1+\alpha^2} + \frac{1}{\alpha}\tan^{-1}\alpha\Big).
\end{equation}
As $\lambda \to 0$, i.e. as the collapse approaches, we learn from (\ref{sna.10}) that
\begin{equation}\label{sna.25}
\frac{w}{v} = 1+\lambda^2 \to 1
\end{equation}
meaning the snake contracts into a sphere. The behavior of $v$ 
follows from (\ref{sna.11}) and (\ref{sna.18})
\begin{equation}\label{sna.26}
v = \frac{W\lambda}{1+\alpha^2\lambda^2}.
\end{equation}
As $\lambda$ goes from 1 to 0, $v$ is a monotonically decreasing function; the snake 
steadily becomes thinner and shorter, before finally vanishing into a point.

\section{The Turtle}\label{S:tur}
For $b=c$, it follows from (\ref{hom.5}) that $v=w$. The system (\ref{hom.7})
reduces to the two differential equations
\begin{equation}\label{tur.1a}
\dot{u} = - \frac{u^2}{v^2}
\end{equation}
and
\begin{equation}\label{tur.1b}
\quad \dot{v} = \frac{u}{v} - 2.
\end{equation}
These equations are identical to the snake equations (\ref{sna.1a}) and (\ref{sna.1b}),
provided we replace $u$ by $w$. The only difference in the solution is:
while $w/v$ was larger than 1 for the snake, the analogous $u/v$ for the turtle
is smaller than 1. The result is that the integral (\ref{sna.7}) is replaced by
\begin{equation}\label{tur.2}
\Big(\frac{u}{w}\Big)^2 + \Big(\frac{u}{v}-1\Big)\Big(\frac{v_1}{v}\Big)^2 = 0
\end{equation}
with the integration constant $v_1$. We use the integral to write $u/v$ as a 
function of $u/v_1$
\begin{equation}\label{tur.3}
\frac{u}{v} = 1 - \Big(\frac{u}{v_1}\Big)^2.
\end{equation}
With (\ref{tur.1a}) we obtain
\begin{equation}\label{tur.4}
\frac{du}{dt} = -\left[1-\Big(\frac{u}{v_1}\Big)^2\right]^2.
\end{equation}
This integrates to
\begin{equation}\label{tur.6}
\frac{1}{v_1}(t_1 - t) = \frac{u/v_1}{2(1-(u/v_1)^2)} + \frac{1}{4}\log\frac{1+u/v_1}{1-u/v_1}
\end{equation}
with constant $t_1$.
For small values of $u/v_1$ we find
\begin{equation}\label{tur.7}
t_1 - t \approx u.
\end{equation}
Since we assume that $u,v>0$ it follows from (\ref{tur.3}) that
\begin{equation}\label{tur.8}
\Big(\frac{u}{v_1}\Big)^2 < 1.
\end{equation}
For
\[
\frac{u}{v_1} \to 1
\]
we obtain
\begin{equation}\label{tur.9}
\frac{1}{v_1}(t_1 - t) \approx  \frac{1}{4(1 - u/v_1)} - \frac{1}{4}\log(1 - u/v_1) + \frac{1}{4}\log 2.
\end{equation}
With these data we are now able to describe the Ricci flow for the turtle. Our initial
conditions are the values of $u$ and $v$ at time $t = 0$. We call these initial
values $U$ and $V$. The ratio $U/V$
\begin{equation}\label{tur.10}
U/V = a(0)/b(0), \qquad b(0) = c(0)
\end{equation}
gives the initial aspect ratio of the turtle. We have
\begin{equation}\label{tur.11}
1 - U/V \equiv \beta^2
\end{equation}
where the parameter $\beta$ measures the degree of initial non-sphericity.
The integral (\ref{tur.3}) allows us to determine the constant $v_1$ that
belongs to the given initial values. We get
\begin{equation}\label{tur.12}
v_1 = \frac{U}{\sqrt{1 - U/V}} = \frac{U}{\beta}.
\end{equation}
We determine the constant $t_1$ by entering the initial values into (\ref{tur.6})
and obtain with (\ref{tur.3})
\begin{equation}\label{tur.13}
t_1 = \frac{U}{2U/V} + \frac{1}{4}\frac{U}{\sqrt{1 - U/V}}\log\frac{1  + \beta}{1  - \beta},
\end{equation}
or
\begin{equation}\label{tur.14}
t_1 = \frac{U}{2(1 - \beta^2)} + \frac{U}{4\beta}\log\frac{1 + \beta}{1 - \beta}.
\end{equation}
We obtain then from (\ref{tur.6}) for the time $t$ as a function of $u$
\begin{equation}\label{tur.15}
t = \frac{U}{2(1 - \beta^2)} + \frac{U}{4\beta}\log\frac{1 + \beta}{1 - \beta}- \frac{u}{2(1-(u/v_1)^2)} - \frac{U}{4\beta}\log\frac{1+u/v_1}{1-u/v_1}.
\end{equation}
Writing
\begin{equation}\label{tur.16}
\mu \equiv \frac{u}{U}
\end{equation}
we have
\begin{equation}\label{tur.17}
t = U\left[\frac{1}{2(1-\beta^2)} - \frac{\mu}{2(1-\beta^2\mu^2)} + \frac{1}{4\beta}\log\frac{(1+\beta)(1-\beta\mu)}{(1-\beta)(1+\beta\mu)}\right].
\end{equation}
At $t = 0$ we have $\mu  = 1$. As $t$ increases, $\mu$ shrinks.
As we see from Fig. \ref{fig.1} and (\ref{sna.x1}) the ratio of $u/v = a/b$ approaches $1$.
This means the contracting turtle becomes relatively thicker before it collapses into a point.
This happens with $\mu = 0$ at time
\begin{equation}\label{tur.18}
T' = \frac{U}{2}\Big(\frac{1}{1-\beta^2} +\frac{1}{2\beta}\log\frac{1+\beta}{1-\beta}\Big).
\end{equation}
We see that the collapse occurs in a finite time.
The behavior of $v = w$ follows from (\ref{tur.3})
\begin{equation}\label{tur.19}
v = w = \frac{U\mu}{1-\beta^2\mu^2}.
\end{equation}
As $\mu$ goes from 1 to 0, $v = w$ is a steadily decreasing function.

\section{The Dragon}\label{S:dra}
For the following, see Fig. \ref{fig.1}. The 
deformation parameters $a, b, c$ are assumed to be in increasing order and their
ratios $y = (b-a)/c$ and $x = (b+a)/c$ are plotted in the $x$-$y$ plane. 
The possible
values form an isosceles right triangle based on the segment of the $x$-axis
from $0$ to $2$. We study the Ricci flow of the Dantes in this diagram.

With initial condition $a = b = c$ at time $t = 0$, the curve of development for the
Ricci flow shrinks into the fixpoint with coordinates $x = 2,y = 0$. In the
first symmetric case, the snake-like configuration, where $a = b$ at $t = 0$,
the initial point lies on the $x$-axis between 0 and 2 and the development 
proceeds then along the $x$-axis, the hypotenuse of the triangle, ending at 
$x = 2$. In the second symmetric case, the turtle-like configuration, where 
$b = c$ at $t = 0$, the development starts on the right cathetus, the right 
edge of the triangle, and proceeds on this line toward the point with coordinates 
$x = 2,y = 0$. Any point inside the triangle is the starting point of a Ricci
flow. We call the shapes interpolating between the snake and the turtle \emph{dragons}.
Since the flow is stationary and scale-invariant, the Ricci flow triangle is
fibered by flow lines. It is these flow lines that we wish to determine.

Their slope $dy/dx$ is given by (\ref{hom.7}) with $R^2 = 4$.
\begin{align}\label{dra.1}
\frac{dy}{dx} &= \frac{d\left[(b-a)/c\right]}{d\left[(b+a)/c\right]} \\
&= \frac{\dot{v} - \dot{u} - \dot{w}(v-u)/w}{\dot{v} + \dot{u} - \dot{w}(v+u)/w} \nonumber \\
&= \frac{[u(u^2-v^2-w^2) - v(v^2-u^2-w^2) + y(2uvw + w(w^2-u^2-v^2))]/w^3}{[-u(u^2-v^2-w^2)-v(v^2-u^2-w^2) - 4uvw + x(2uvw + w(w^2-u^2-v^2))]/w^3} \nonumber .
\end{align}

With 
\begin{equation}\label{dra.2}
\frac{u}{w} = \frac{1}{2}(x-y), \quad \frac{v}{w} = \frac{1}{2}(x+y) 
\end{equation}
this gives
\begin{equation}\label{dra.3}
\frac{dy}{dx} = \frac{y(x^2+y^2-2)}{y^2(2x-1)+x(x-2)}.
\end{equation}
The three linear functions
\begin{equation}\label{dra.4}
y = 0, \quad y = 2 - x, \quad y = x
\end{equation}
solve equation (\ref{dra.3}). They form the sides of the triangle in Fig. \ref{fig.1}.
The numerical integration shows the development of 
the dragons. All curves emerge from the origin and end at $x = 2,y = 0$. They 
reach their maximum on a circle of radius $\sqrt{2}$ about the origin. Near the
origin the dragons emerge as very thin hardly flattened snakes. Under the Ricci
flow they all collapse spherically into points.

\begin{figure}[htp]
\centering
\includegraphics[width=140mm]{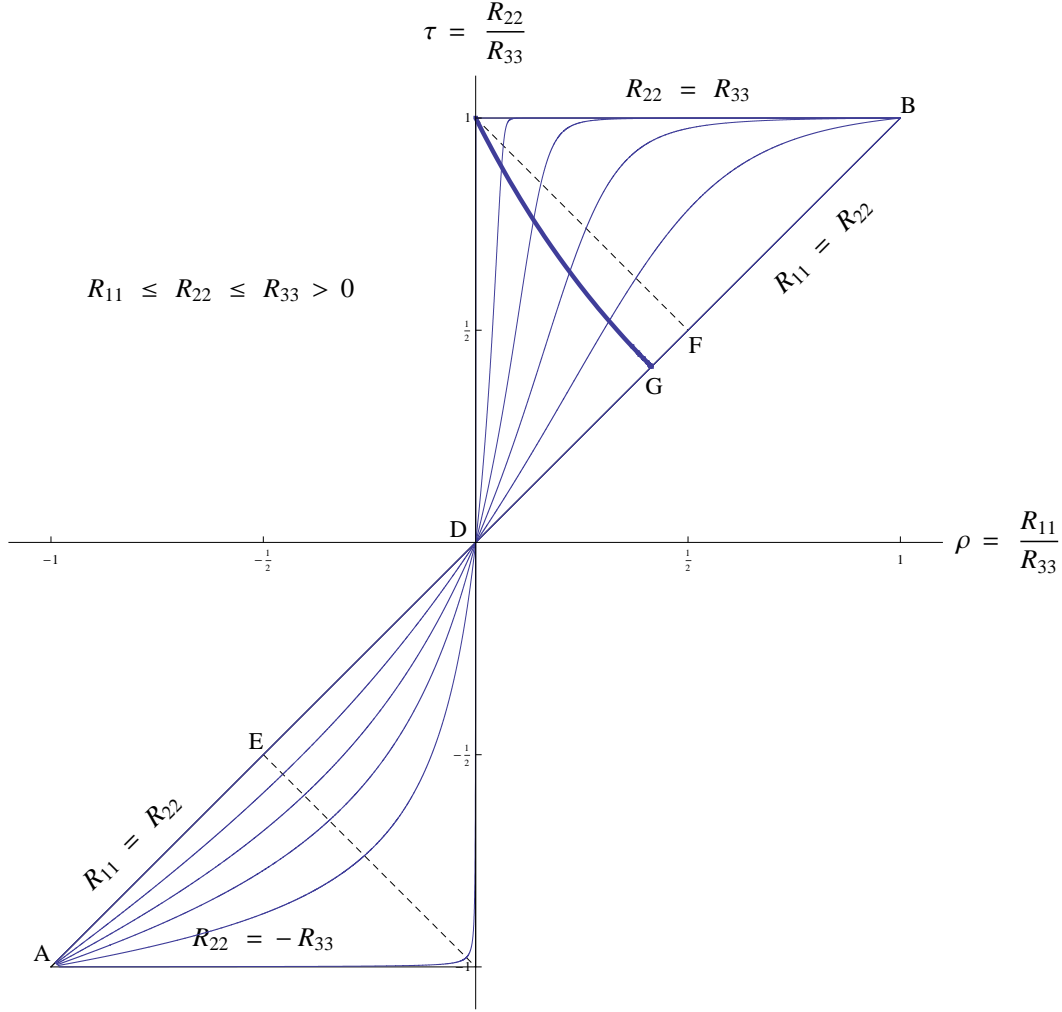}
\caption{\textbf{Diagram for the eigenvalue ratios of the Ricci tensor in a stretched $S^3$}.
\newline
Point $B$ denotes again the isotropically stretched $S^3$.
Here $R_{11} = R_{22} = R_{33}$, the eigenvalues of the Ricci tensor are equal and so are the principal curvatures.
The segment from $-1$ to $1$ of the $R_{22}/R_{33}$-axis, except the origin $D$, does not represent possible configurations.
While the map from Fig. \ref{fig.1} to this figure is one-to-one with corresponding points denoted by the same letters, the line $CD$ in Fig. \ref{fig.1} collapses here into the point $D$.
If $R_{11} = 0$, then $R_{22}$ has to vanish too. }\label{fig.2}
\end{figure}	

We turn now to Fig. \ref{fig.2}, which pictures the ratios of the eigenvalues of the
Ricci tensor. We define
\begin{equation}\label{dra.5}
R_{22}/R_{33} = \rho, \quad R_{11}/R_{33} = \tau
\end{equation}
and obtain from (\ref{ric.18})
\begin{equation}\label{dra.6}
\rho = \frac{a + b - c}{a + c - b} = \frac{u/w + v/w - 1}{u/w + 1 - v/w} = \frac{x - 1}{1 - y},
\end{equation}
\begin{equation}\label{dra.7}
\tau = \frac{a + b - c}{b + c - a} = \frac{u/w + v/w - 1}{v/w + 1 - u/w} = \frac{x - 1}{1 + y}.
\end{equation}
This transformation allows us to transfer the Ricci flow lines from Fig. \ref{fig.1} to
Fig. \ref{fig.2}. Since the Ricci flow lines are not symmetric with respect to reflection
at the line $x = 1$ (the line CD in Fig. \ref{fig.1}) the flow is also not symmetric with
respect to reflection on the origin in Fig. \ref{fig.2}. All flow lines appear to start
in point A and end in point B. All lines go through the origin of Fig. \ref{fig.2}. The
eigenvalue  0 of the Ricci tensor is degenerate here. That means, according to
(\ref{ric.18}),
\begin{equation}\label{dra.9}
\kappa_1 = \kappa_2 = \frac{c^2}{R^2}, \quad \kappa_3 = -\frac{c^2}{R^2}.
\end{equation}
It would be nice if one could visualize this very peculiar snake.

\noindent {\it Acknowledgments}~ We are very grateful to Prof. Peter Ozsv\'ath for providing help with the figures based on his work with Mathematica. I.O. thanks The University of Texas at Dallas for support.

\end{document}